  \documentstyle[prd, preprint,eqsecnum,aps]{revtex}  
  \begin{document}
  \draft
   \preprint{\vbox{
   \hbox{UM-TH-99-12,SINP-TNP/99-39}
   \hbox{hep-th/9912165}
  }}

  \title{$(p + 1)$-Dimensional Noncommutative Yang-Mills\\
 and D(p$-$2) Branes\\ }

  \author {J. X. Lu$^1$ and Shibaji Roy$^2$}
  \address{$^1$Randall Physics Laboratory,
  University of Michigan, Ann Arbor, MI 48109-1120\\
  $^2$Saha Institute of Nuclear Physics,
  1/AF Bidhannagar, Calcutta 700 064, India\\
   E-mails: jxlu@umich.edu, roy@tnp.saha.ernet.in}

  \maketitle
  
  \begin{abstract}
  We consider systems of non-threshold bound states (D(p$-$2), Dp), 
  for $2\le p \le 6$, in type II string theories.
  Each of them can be viewed as Dp branes with a nonzero (rank
  two) Neveu-Schwarz $B$ field. We study 
  the noncommutative effects in the gravity dual descriptions of 
  noncommutative 
  gauge theories for these systems in the limit where the brane worldvolume 
  theories decouple from gravity. We find that the noncommutative 
  effects are actually due to the presence of infinitely many D(p$-$2) branes
  in the (D(p$-$2), Dp) system which play the dominant role over the Dp branes
  in the large $B$-field limit. Our study indicates that Dp branes with
  a constant $B$-field represents dynamically the system of infinitely
  many D(p$-$2) branes  without $B$-field in the decoupling limit.
   This implies an
  equivalence between the noncommutative Yang-Mills in $(p +
  1)$-dimensions and an ordinary Yang-Mills with gauge group $U (\infty)$
  in $(p - 1)$-dimensions. We provide a physical explanation for the new
  scale which measures the noncommutativity. 
  \end{abstract}
  \newpage

  \section{Introduction\protect\\}
  \label{sec:intro}
  
  Gauge theories on noncommutative spaces can arise in certain limits of 
  string theory\cite{how,cds,douh,ardas,bigs}. For a system of Dp branes, a 
  constant Neveu-Schwarz (NS) $B$ field in spatial directions on the 
  worldvolume is the key to making the space
  noncommutative. Following Maldacena's AdS/CFT correspondence for the usual
  Yang-Mills theory, it is quite
  natural to look for the gravity duals of the noncommutative gauge theories 
 (NCYM).
  In order to do so, one has to identify a limit in which 
  closed strings decouple 
  from the open strings which end on the Dp-branes and the dynamics for
  the open
  strings is entirely described by their low energy gauge theory on a 
  noncommutative space\footnote{We will refer to this limit as the NCYM
  decoupling limit as opposed to the usual decoupling limit for the AdS/CFT
  correspondence of Maldacena.}. Recently, this has been achieved in general by
  Seiberg and Witten \cite{seiw}, and for special systems by 
  Hashimoto and 
  Itzhaki \cite{hasi} and by Maldacena and Russo \cite{malr}. 
  The correspondence between NCYM and the gravity duals is proposed in 
  \cite{hasi,malr}. This correspondence has also been studied for many
  other known systems recently in \cite{aliosj,haro}. The thermodynamics
  for NCYM has also been studied in \cite{malr,barr,caio,haro}. 
  Seiberg and Witten have shown that 
  noncommutative gauge theories are equivalent to usual gauge theories by a 
  change of variables\footnote{This in general does not require the two gauge
  groups in the respective descriptions to be the same. Rather we only need
  to preserve the gauge equivalence relation. In other words, whenever
  two ordinary gauge fields are gauge-equivalent by an ordianry gauge
  transformation, we have two corresponding gauge-equivalent noncommutative
  gauge fields which are related by a noncommutative gauge
  transformation.
  However, in this paper, when we talk about `gauge group', we always
  mean the unitary group $U(n)$ associated with the number of D-branes 
  defined in terms of $n \times n$ hermitian matrices, not the one
  associated with the noncommutative, associative algebra defined in terms of
  $\ast$ product for functions, i.e., 
  $f \ast g = f g + (1/2) i \theta^{ij} \partial_i f \partial_j g + {\cal O}
  (\theta^2)$. }. 
  This can be understood physically by the fact that 
  the two descriptions are due to the different choices of regulator in the 
  sigma model expansion and physics in a well-defined theory should be 
  independent of such choices. 

      Whether we have the usual YM or NCYM depends crucially on the
vanishing or the nonvanishing of the constant worldvolume 2-form  
field\footnote{The constant $B$-field can be taken as the asymptotic
  value of the NS $B$-field in a gravity configuration. Note that our NS
  2-form $B$ differs from that of Seiberg and Witten by a 
factor $2 \pi \alpha'$,
  i.e., $B = 2\pi \alpha' B^{\rm SW}$.} 
${\cal F} \equiv 2 \pi \alpha' F + B$. 
  Here $B$ is the NS 2-form field
  and $F$ is the worldvolume 2-form gauge field strength. 
  So whenever we have a constant ${\cal F}$-field, we can potentially
have a NCYM description. For
  convenience, we, as usual, make a
  gauge choice such that we have $F = 0$ while $B$ takes the given
  constant value of ${\cal F}$. 

  It is well-known that the constant $F$ field on the worldvolume of a Dp
  brane implies that
  there are infinitely long fundamental strings 
(for short, F-strings)
  and/or D(p$-$2) 
  branes in the Dp-brane from the spacetime perspective 
  \cite{bremm,arfsj,lurone,lurtwo,lurthree} (actually there are infinitely
  many such strings or D(p$-$2) branes). The former gives the so-called 
  (F, Dp) non-threshold
  bound states \cite{arfsj,lurone,lurtwo} while the latter gives
 (D(p$-$2), Dp) non-threshold bound states 
  \cite{bremm,lurone,lurthree}\footnote{There actually exist 
\cite{lurthree,lurfour} more
  complicated configurations of Dp-branes with non-vanishing NS $B$-field.  
  For example, the so-called ((F, D1), D3) non-threshold bound
  state given in \cite{lurthree} has been used to discuss the gravity
  dual of NCYM in \cite{malr}.}.
  This relationship between the constant $F$-field
on the worldvolume of a Dp-brane and the
  F-strings and/or D(p$-$2)-branes living on a Dp-brane  must
  imply a close connection between the non-commutativity and these
  F-strings and/or D(p$-$2) branes.

  The aim of this paper is to reveal the connection and possible 
  consequences of this. 
  In general, if there exists a non-threshold BPS bound
  state\footnote{In this paper, we always consider infinitely extended
  rather than compactified branes. One advantage in choosing so is to
  avoid considerations of possible finite size effects as discussed in 
  \cite{bigs} and non-local light winding modes as
  discussed in \cite{douh,hasi,hasione}. } consisting
  of $p'$-branes and $p$-branes with $p' < p$,   the charge quantizations 
  imply that $2\pi \alpha'F + B$ is also quantized. It can actually be 
  determined in terms of the quantized (integral) charges $q$ and $n$ 
   (see \cite{bremm}, for example\footnote{It
  can also be inferred from the contribution of  $2\pi \alpha'F + B$ to
  the tension of the corresponding non-threshold bound state as given 
  in \cite{bremm,arfsj,lurone,lurtwo,lurthree}.}), where the integer $n$ is
  the number of $p$-branes and the integer $q$ is the number of 
  parallel $p'$-branes  per $(2 \pi)^{(p - p')} \alpha'^{(p -p')/2}$ area
  over the $(p - p')$-plane perpendicular to the $p'$-branes in the
  worldvolume of $p$-branes. We therefore expect that the
  non-commutativity is actually related to the integral charge $q$. 
  As we will see, this is indeed true.

  For concreteness, we will focus, in this paper, on the systems of 
  non-threshold bound states (D(p$-$2), Dp), for $2 \le p \le 6$, in type
II string theories. The gravity configurations of
  these  bound states have been given in 
  \cite{bremm,tser,lurthree}. The asymptotic value of $B$-field always
  vanishes for each of these gravity solutions\footnote{This is also true 
  for all known gravity solutions of Dp branes with non-vanishing
  $B$-field,  see for examples
  \cite{bremm,tser,lurtwo,lurthree,lurfour}.}. 
  As mentioned earlier, we choose to work in the gauge where the
  constant flux of the worldvolume gauge field strength vanishes. By doing so,
 we
  end up with a non-vanishing asymptotic value for the $B$-field in each
  of the gravity solutions. The component of this asymptotic $B$-field, 
  denoted as $B_\infty$, is actually given as\footnote{In general, we
  could have $B_\infty = a (q/n)$ with $a$ an arbitrary constant. For
  the reasons given in footnote 5, we always keep this dimensionless
  parameter $a$ fixed even in the NCYM decoupling limit. As discussed in
  detail in the following section, our philosophy of NCYM decoupling
  limit is to treat the coordinates in the two co-dimensions of the D(p
  - 2) and Dp branes in (D(p - 2), Dp) similarly as the radial
  coordinate for the bound state in the decoupling limit. The former
  decouples any vortex of q D(p - 2) branes  from the rest in the two
  co-dimensions while the latter decouples the asymptotic region from
  the near-horizon region. This leads to $q \gg n$ in the NCYM decoupling
  limit. However, if the two co-dimensions form a torus rather than an
  infinite plane (therefore, we may have possible complications as mentioned
  in footnote 5), the dimensionless parameter $a$ will be 
$\sim {\tilde b}^2/({\tilde V}_2 \alpha')$ with ${\tilde V}_2$ the area
  for the scaled torus in the decoupling limit. Therefore, $q/n \sim
 {\tilde V}_2 /{\tilde b}$ which appears to be finite in the decoupling
  limit. This latter case will be discussed in \cite{caioone}. Apart from the 
apparent difference for $q/n$ in the decoupling limit, the rest
  conclusions drawn in the two cases are essentially the same. We are 
grateful to R.-G. Cai and N. Ohto for their question which leads
  us to the above discussion.
 Actually,
the above scaled area ${\tilde V}_2$ for the torus is not quite
  independent of the noncommutative parameter $\tilde b$. If we denote
the two coordinates as ${\tilde x}_{p - 1}, {\tilde x}_p$ (we discuss
 them in the following section), we have the noncommutative relation
as $[{\tilde x}_{p - 1}, {\tilde x}_p] = i \tilde b$. So we should have
${\tilde V}_2 \ge \tilde b$. Therefore, we have $q \ge n$. This implies
  that whenever we have large $n$, we must have large $q$ but not
the other way
  around.}
   $B_\infty  = \tan  \theta = q/n$, i.e.,
  quantized in terms of the integer $q$ and $n$ described
  above\footnote{The integer $q$ resembles, to some extent to the 
integer $n$. 
 As discussed in \cite{bremm,lurtwo}, if one calculates the  Noether
  charge associated with the infinitely many D(p$-$2) branes in 
  (D(p$-$2), Dp), the charge
  itself becomes infinity. But the charge over the 2-dimensional
  area $(2\pi)^2\alpha'$ perpendicular to the D(p$-$2) branes in the
  Dp brane 
  worldvolume is
  finite and is given as $(q /(2\pi)^2 \alpha') Q_0^{p -
  2}$ with 
  $Q_0^{p - 2} = (2\pi)^{(11 - 2p)/2} \alpha'^{(5 - p)/2}$ where the
   conventions for the charge is given in \cite{dufkl}. But 
  $Q_0^{p - 2}/((2\pi)^2 \alpha')$ is just the 
  charge units for the $(p + 2)$-form charge associated with the Dp
  branes. So this charge density looks like $q$ units of Dp brane charge.
   Further, the tension for (D(p$-$2), Dp) in string metric is
  $T (D(p - 2), Dp) = \frac{T_0^p}{g_s} \sqrt{q^2 + n^2}$
  with $T_0^p = 1/[(2\pi)^p \alpha'^{(p + 1)/2}]$  the Dp-brane
  tension units
  and $g_s$ the asymptotic string coupling. In the above tension formula,
since the integers
  $q$ and $n$ appear symmetrically, so, in some sense, the
  D(p$-$2) branes in (D(p$-$2), Dp) appears as if they are $q$
  Dp-branes.}.

 In the NCYM decoupling limit, the large $B_\infty$ is the key for the 
origin of NCYM. More
  precisely, as shown in
  \cite{hasi,malr,aliosj}, in order to obtain NCYM we need 
$B_\infty = \tan \theta = 
  {\tilde b}/\alpha'$ with $\alpha' \rightarrow 0$ and $\tilde b$ = fixed.
  For fixed $B_\infty$,
  we end up only with the usual YM as demonstrated in \cite{lurthree}
  and also pointed out in \cite{malr}.  However, so far only 
  the classical $B_\infty = \tan \theta$ has been used in 
  discussing NCYM \cite{hasi,malr,seiw,liw,aliosj}. But as we have
  discussed above,  $B_\infty = \tan \theta$ is in fact quantized 
according
  to $B_\infty = \tan \theta = q/n$. So the above NCYM decoupling limit 
  implies $q/n = {\tilde b}/\alpha' \rightarrow \infty$ in the low 
  energy limit $\alpha' \rightarrow 0$. In other words, we have 
  $q \gg n$ in the NCYM decoupling limit\footnote{see the explanation
given in footnote 8}.    
  One of our observations in this paper is that: 
  {\it in the NCYM  decoupling limit, the infinitely many 
  D(p$-$2) branes in the (D(p$-$2), Dp) system plays dominating role
  over the  $n$ Dp branes in the system. The non-commutativity actually
  arises due to these D(p$-$2) branes and the $(p + 1)$-dimensional NCYM
  is just one way of describing these D(p$-$2) branes.}   
 
	Some closely related work and ideas have been presented some
time ago by a number of authors. 
 Townsend first suggested an equivalence  
 between a D2 brane and a condensate of D0 branes
  \cite{tow}.
  His proposal was
  based on the following observation: A 
  condensate of D0 branes can be described effectively by
  one-dimensional super YM theory with gauge group $U (\infty)$, i.e.,
  $U(n)$ supersymmetric gauge quantum mechanics (SGQM) with $n
  \rightarrow \infty$. But this is just another description of D2-brane
  which can be inferred from the description of $D = 11$ closed supermembrane
  by a $U(n)$ SGQM for $n \rightarrow \infty$ established in \cite{dewhn} 
  in the old days of  $D = 11$ supermembrane theory. This is also one of
  the motivations for the original Matrix theory proposal of M-theory
  \cite{banfss}. Generalizing this to other D-branes, Ishibashi argued
  that a D-brane in a constant $B$-field background can be described 
  equivalently as a collection of infinitely many lower dimensional
  D-branes \cite{ishone}. However, the validity of the above equivalence is
  not discussed in \cite{ishone}. It appears that this has been remedied
  in \cite{cors}. The authors in the latter paper argued that 
  a D2 brane with a constant worldvolume gauge field strength in the 
  Sen-Seiberg limit \cite{sen,sei} is described by the Matrix theory quantum
  mechanics action of D0 branes.    

	Given the requirement of constant $B$-field background for the
above equivalence, it is natural for these authors to seek the connection
of their work to the NCYM description of D branes with this background.
They find whether one has an ordinary or noncommutative description
depends on the gauge choice of the D brane worldvolume diffeomorphism 
\cite{cors,ishtwo}. If one takes the static gauge, i.e., the coordinates
 parallel to the brane are fixed and the worldvolume gauge field remains
dynamical, one ends up with an ordinary Yang-Mills description. On the
other hand, if one chooses the so-called ``constant field strength
gauge'', i.e., no fluctuations of the ordinary gauge field are allowed,
then one ends up with NCYM description. In this latter gauge, the
dynamical degrees of freedom, which are usually described by the ordinary
worldvolume gauge field, are now carried by the scalar fields
corresponding  to the parallel coordinates of the brane. The
noncommutative gauge field appears as the fluctuations of these scalar
fields with respect to the static BPS configuration\cite{ishtwo,cor}.
Especially in \cite{cor}, the author pointed out that the decoupling
limit for NCYM corresponds to the Sen-Seiberg limit. The two
descriptions can be mapped to each other through the worldvolume 
diffeomorphism \cite{ishtwo,cor,oku}.    

	One of the goals in the present paper is to show that in the
gravity dual description, the $(p + 1)$-dimensional $U (n)$ NCYM is equivalent 
to an ordinary $(p - 1)$-dimensinal YM with gauge group $U(q)$ with
$q/n = \tilde b /\alpha' \rightarrow \infty$ and with $\tilde b$ as the 
noncommutative parameter. In
other words, $n$ Dp branes with  $B_{\infty} = q/n$ is
equivalent to infinitely many D(p$-$2) branes without $B$-field in the 
NCYM decoupling
limit. Therefore, this work lends support to what is just described above.
	        
 Let us now
  examine the relation $q/n = {\tilde b}/\alpha'$ more closely. 
  We know that the
  fixed parameter $\tilde b$ is directly related to the non-commutative
  geometry. The larger (smaller) the ${\tilde b}$ is, the more (less) 
  non-commutative the space geometry will be. The presence of this 
 dimensionful parameter implies that a new scale is introduced into the 
 theory under consideration. We will spell out the meaning of this new
 scale in the following section. To make sense of the
  system (D(p$-$2), Dp), we need $n \ge 1$. Large $\tilde b$ requires
  large $q$ and small $n$ from the relation $q/n = \tilde b /\alpha'$. 
  So for NCYM we actually need large $q$
  and small $n$ rather than large $n$. In other words, large $n$ really
  goes against the NCYM
  effect even though it is good for the gravity dual description in the
  usual AdS/CFT correspondence. It seems as if there were an exception,
  i.e., we could take $n$ large with a fixed $\tilde b$. 
  But from $q/n = \tilde b/\alpha'$, we see that we can increase $n$ 
only at the
  price of increasing $q$ for a fixed $\tilde b$. For example, for a
  fixed $\tilde b$, when $n = 1$, we denote $q = q_0$ which gives $q_0 =
  \tilde b /\alpha'$. Now we want to increase $n = 1$ to a large $n$
  while keeping $\tilde b$ unchanged. Then we must have $q = n q_0$.
  
  Because the new parameter $q$ or
  $\tilde b$ accounts for the non-commutative effect in NCYM and if we
  really want to focus on this effect, one should take large 
$q$ (or large
  $\tilde b$) and small $n$ even in the gravity dual description of NCYM
  which is contrary to what have been claimed in the
  literature. Interpreted in another way, the real meaning for
  ``large $n$" used in the literature is the large $q$.
   
  So in order to isolate the non-commutative effect and to find
  the reason behind this effect, we should limit ourselves to small
  $n$. Without the loss of generality, we can simply set $n = 1$ from now
  on\footnote{When we take $n = 1$, we have the `gauge group' $U (1)$
  for NCYM. For a general $n$, we have the `gauge group' $U(n)$.}.
   As we will demonstrate in the following sections using the gravity
  duals of NCYM, whenever the gravity dual description is valid, the
  (D(p$-$2), Dp) system is  reduced to  a system of
  infinitely many D(p$-$2) branes with
  two additional isometries and without $B$-field.  
        
	This paper is organized as follows. We will present the general 
features valid for all (D(p$-$2), Dp)
   systems with $2 \le p \le 6$ in the following section. In section 3, 
  we will provide detail evidence in supporting our above claims for
  the well-studied system (D1, D3). In section 4, we will discuss similar
 evidence for each of the remaining systems. We conclude this paper 
  in section 5.

\section{General Properties}

     To begin with let us explain the new scale $\tilde b$ further.
In terms of the $(p + 1)$-dimensional NCYM description, this new scale
defines the noncommutativity. In the corresponding ordinary 
$(p + 1)$-dimensional Yang-Mills description, this new scale appears as
a dimensionful coupling constant in addition to the usual gauge
coupling, therefore giving rise to higher dimensional operators.
All these are due to the constant background $B$-field. We know that
the relevant energy scale is $u$ (i.e., the energy carried by an open string 
stretched between a probe and the bound state which will be defined
as $r = \alpha' u$ with $r$ the radial coordinate shortly). 
We also know that the physical effect of noncommutativity is not solely
determined by $\tilde b$ but by some combination of $\tilde b$ and
$u$, i.e., $\sim {\tilde b}^{1/2} u$. For example, even for large $\tilde b$,
if ${\tilde b}^{1/2} u \ll 1$ the noncommutative effect is still
negligible as pointed out in \cite{malr} (One may naively think that
large noncommutativity, i.e., large $\tilde b$, implies its large
physical effect). It is hard to understand this if one sticks with the
either noncommutative or ordinary Yang-Mills description of Dp branes.

	As we have discussed in the previous section and will
demonstrate in this paper, in the NCYM decoupling limit,
the (D(p$-$2), Dp) system is reduced to that of infinitely many 
D(p$-$2) branes with two additional isometries and without $B$-field. It is
much easier to understand the above in the ordinary Yang-Mills 
description of these infinitely many D(p$-$2) branes. Unlike the Dp
branes, there are two kinds of transverse spaces to the D(p$-$2)
branes. One is the common transverse space shared by both of the 
D(p$-$2) and Dp branes in (D(p$-$2), Dp). The corresponding field theories 
share the common energy scale $u$. Now the D(p$-$2) branes have
another transverse space with co-dimension two with respect to the Dp
brane worldvolume. Since this space is treated differently from the
previous one, there must be another energy scale for the D(p$-$2)
branes. In other words, the field theory describing the low energy
dynamics of the infinitely many D(p$-$2) branes needs two energy
scales one of which is $u$. What is the second energy scale?

We can estimate this scale in two different ways. For convenience, let,
say, the D(p - 2) branes lie along $x^0, x^1, \cdots, x^{p - 2}$
directions and the Dp brane along $x^0, x^1, \cdots, x^{p - 2}, x^{p -
1}, x^p$ directions. As discussed in the previous section, 
we have $q$ D(p$-$2) branes per $(2 \pi)^2 \alpha'$ area over the
infinitely extended $x^{p - 1} x^p$-plane. We can also view each set of
q D(p - 2) branes through the $x^{p - 1} x^p$-plane as a vortex on the
plane. From the worldvolume viewpoint, these vortices are magnetic ones.
In some sense, they resemble the vortices in type II superconductivity.
Each of these vortices has an area on the order of $\alpha'$ and the
distance between the centers of two nearby vortices is on the order
$\sqrt{\alpha'}$. Now we can estimate the characteristic separation $d$ 
between two such D(p$-$2) branes in a given vortex as   
\begin{equation}
d = \left(\frac{(2 \pi)^2 \alpha'}{q}\right)^{1/2}.
\label{eq:dis}
\end{equation}
If we substitute $q = \tilde{b}/\alpha'$ obtained in the previous
section in the above, we have 
$d \sim \alpha'/{\tilde b}^{1/2}$. The energy carried by an open string 
stretched between two such D($p - 2$) branes in a given vortex 
is given as $ d/\alpha'
\sim 1/{\tilde b}^{1/2}$ which should define the second energy scale.

As we will discuss latter in this section, the NCYM decoupling limit
also consists of
$\alpha' \rightarrow 0, r = \alpha' u, x_{(p - 1), p} = 
(\alpha'/{\tilde b})\, {\tilde x}_{(p - 1), p}$ with $u$ and ${\tilde
x}_{(p - 1), p}$ fixed. The limit $\alpha'
\rightarrow 0$ means a low energy field theory limit.
$r = \alpha' u$ means that in the gravity description, only the 
near-horizon region is relevant and it also defines the first energy
scale $r/\alpha' = u$ mentioned above. In analogy to $r = \alpha' u$,
$x_p = (\alpha'/{\tilde b})\, {\tilde x}_p$ (the same for $x_{p - 1}$)
defines the distance between a dynamical probe and the vortex of q 
D(p - 2) branes located at ${\tilde x}_p ={\tilde x}_{p - 1} = 0$. Hence the $x_p$ is
the length of an open string stretched between the probe and the
vortex. Therefore, the
second energy scale should also be given as 
$x_p /\alpha' = {\tilde x}_p/\tilde b$ which is finite. Note, however,
that the open
string stretched between two nearby vortices carries an energy on the 
order of $\sqrt{\alpha'}/\alpha' = 1/\sqrt{\alpha'} \rightarrow \infty$ 
as $\alpha' \rightarrow 0$ since the open string length is on the order
$\sqrt{\alpha'}$. In other words, the nearby vortices have
no dynamical influence on the vortex under consideration in the NCYM
decoupling limit. So we need to consider only one vortex and the rest 
vortices decouple from this one. 

	It appears in the above that we obtain two different estimations 
for the second energy scale. Let us resolve this apparent difference.
The second approach gives this scale as ${\tilde x}_p /\tilde b$. The
${\tilde x}_p$ should be viewed as the quantum-mechanical
 average of the corresponding operator. Because of the 
noncommutative relation $[\tilde x^{p - 1}, \tilde x^p] = i  \tilde b$
for the corresponding operators, we should have the value for both of
${\tilde x}^{p - 1}$ and $\tilde x^p$ on the order of ${\tilde b}^{1/2}$.
Therefore, we have the energy scale on the order $1/{\tilde b}^{1/2}$
which agrees with what we have obtained from the first approach.           
So we explain why we have a second energy scale and how this energy
scale is related to the parameter of noncommutativity $\tilde b$. 

	Because of the presence of two energy scales, i.e., 
$u$ and $1/{\tilde b}^{1/2}$, we have
opportunities to examine the low energy dynamics for these D(p$-$2) branes.
As is understood, the lower scale is relevant for the low energy
dynamics. If $1 /{\tilde b}^{1/2} \ll u$, i.e., ${\tilde b}^{1/2} u \gg
1$, the dynamical degrees of freedom associated with the scale 
$1/{\tilde b}^{1/2}$ are dominantly important while those associated with
the scale $u$ are not. Roughly speaking, the ordinary ($p -
1$)-dimensional Yang-Mills with gauge group $U (q)$ (with 
$q \rightarrow  \infty$) explores
only $p + 1$ spacetime dimensions dynamically. Since the scale
$1/{\tilde b}^{1/2}$ is associated with open strings stretched between
the D(p$-$2) branes rather than Dp branes, the natural 
description of the dynamics is in terms of these D(p$-$2) branes even
though it may not be the best.
However, if we force ourselves to describe the dynamics in terms of Dp
branes instead, we have to do something unnatural. This unnatural twist
is to make the two codimensions noncommutative and we expect that the
effect of noncommutativity is important. This is indeed true. As
we will demonstrate in the following sections. The validity of the
gravity dual description of NCYM requires ${\tilde b}^{1/2} u \gg 1$.
Further the gravity configuration of Dp brane with a constant $B$-field is 
reduced to that of D(p$-$2) branes with two isometries and without
$B$-field. The importance of the noncommutative effect can also be seen
from the noncommutative $\ast$ product
for functions $f \ast g = f g + (1/2) i \theta^{ij} \partial_i f
\partial_j g + \cdots$. If we choose to describe the underlying dynamics
in terms of Dp branes, the apparent energy scale is $u$. Therefore, 
$\partial f \sim u f$. Since $\theta^{ij} \sim \tilde b$ for the 
present case,        
we see that the leading noncommutative modification to the usual
function product is of the order $\tilde b u^2 fg$. In the present case,
i.e., for ${\tilde b}^{1/2} u \gg 1$, the noncommutative effect is indeed 
important as anticipated.

	On the other hand, if $1/{\tilde b}^{1/2} \gg u$, i.e., 
${\tilde b}^{1/2} u \ll 1$, then only the dynamical degrees of freedom
associated with the energy scale $u$ are important. In this case, the
appearance of the infinite 
D(p$-$2) branes only collectively modifies the ($n = 1$) Dp brane
without B-field through the correction of order ${\tilde b}^{1/2} u \ll
1$.
In other words, we do not expect the corresponding noncommutative effect to be 
important if we choose to describe the underlying dynamics in terms of
the Dp brane instead. This can be easily seen from the above $\ast$ product of 
functions. The gravity description of the NCYM is not valid in this
region. However, if we are willing to extrapolate to this region as done 
in \cite{malr}, the resulting gravity description is nothing but the 
 gravity dual of the ordinary Yang-Mills for Dp branes without
$B$-field. The above can also be interpreted as the usual large $n$
limit. We first set $q = {\tilde b'}/\alpha'$ with a finite $\tilde b'$.
Then $q/n = (\tilde b'/n)/\alpha'$. We take $\tilde b = \tilde b'/n$. So
in the large $n$ limit, the U(n) NCYM is essentially the same as the
corresponding ordinary U(n) YM. In other words, with fixed $q$ behavior
(i.e., fixed $\tilde b'$), in the large n limit, the U(n) NCYM becomes 
an ordinary U(n) YM.

	   Let us provide further evidence in supporting our above 
explanation.  In the presence of constant $B^{\rm SW}$-field, Seiberg
and Witten have shown in \cite{seiw} that the metric seen by the open
strings on the Dp-branes is the effective open string metric 
$G^{\rm SW}_{MN}$ rather than the closed string one 
$g^{\rm SW}_{MN}$ ($M, N = 0, 1,
\cdots, 9$). In the NCYM decoupling limit, i.e., with $G^{\rm
SW}_{MN}$ and $B^{\rm SW}$ fixed and $\alpha' \rightarrow 0$, 
$g^{\rm SW}_{MN}$ should vanish according to $g^{\rm SW}_{MN} \sim
\alpha'$ along the directions in which $B^{\rm SW}$ are non-zero. 
They also give the relation between the 
open string coupling $G_s$ and the closed string coupling $g_s$ as
\begin{equation}
G_s = g_s \left(\frac{{\rm det} \left(g^{\rm SW} + 2 \pi \alpha' B^{\rm
SW}\right)}{{\rm
det} \,g^{\rm SW}}\right)^{1/2},
\label{eq:occr}
\end{equation}
where the metric $g^{\rm SW}$ and $B^{\rm SW}$ take values only along the
Dp-brane directions. The low energy Yang-Mills coupling is given in
terms of the open string coupling $G_s$ as
\begin{equation}
g^2_{\rm YM} = (2 \pi)^{p - 2} \alpha'^{(p - 3)/2} G_s.
\label{eq:ymc}
\end{equation}
As pointed out in \cite{seiw}, the $G_s$ and $g_s$ must scale in the
following ways
\begin{equation}
\frac{G_s}{(\alpha')^{(3 - p)/2}} = {\rm fixed},\qquad
\frac{g_s}{\alpha'^{(3 - p + r)/2}} = {\rm fixed},
\label{eq:scaling}
\end{equation}
such that $g^2_{\rm YM}$ can be kept finite for a quantum theory 
in the decoupling limit. We note that the scaling of the closed string
coupling depends on the rank $r$ of the $B$-field.    

	 For our (D(p$-$2), Dp) systems, we always have $r = 2$. 
So $g_s = \alpha'^{(5 - p)/2} \tilde {g} = \alpha'^{[3 - (p -
2)]/2}\tilde g$ with 
$\tilde g$ fixed in the decoupling limit. If we look from the gravity 
dual viewpoint of the NCYM,  only the closed string coupling 
$g_s$ is relevant. Then the above scaling behavior of $g_s$ is 
actually the same as that for simple D(p$-$2) branes (i.e., without
$B$-field) rather
than simple Dp branes 
in the decoupling limit for the usual correspondence discussed in 
\cite{itzmsy}. 
For example, for $p = 5$, i.e., for (D3, D5) system, the $g_s$ itself
is fixed in the NCYM decoupling limit which is the same as that for the 
 simple D3 rather than the simple D5 branes in the usual decoupling
limit. This is entirely consistent with our above explanation. Since the 
gravity description is classical, we do not sense noncommutativity of the 
spatial directions along which constant $B$-field does not vanish. Then our
above discussion implies that this description must be that of
infinitely many (i.e., q) D(p$-$2) branes 
without $B$-field. This is indeed true as we will demonstrate. This
explains the behavior of the closed string coupling in the decoupling
limit.
    
However, if we look from
the open string viewpoint, the open string coupling $G_s$ is relevant.
From Eq.\ (\ref{eq:scaling}), we have $G_s = \alpha'^{(3 - p)/2}
 {\tilde g} \tilde b$, which will be derived in the
following. Comparing this scaling behavior with what has been given in
\cite{itzmsy} for simple Dp branes in the decoupling limit for the
usual YM, one can see that $G_s$ scales like that for simple Dp branes rather
than for simple D(p$-$2) branes as anticipated.

Now let us study the gravity configurations for (D(p$-$2), Dp) systems
and for D(p$-$2) branes with two additional isometries in the NCYM decoupling
limit, respectively. 

The string-frame metric, the dilaton and  the $B$-field for (D(p$-$2),
 Dp) with $2 \le p \le 6$, first given in \cite{bremm}, can be expressed
 in a unified way as
\begin{eqnarray}
d s^2 &=& H^{1/2}\left[H^{-1} \left (- d x_0^2 + d x_1^2 + \cdots + 
d x_{p - 2}^2 \right) + H'^{-1} \left( d x_{p - 1}^2 + d x_p^2\right) 
+ d y^i d y^i \right],\nonumber\\
e^{2 \phi} &=& g_s^2 \,\frac{H^{(5 - p)/2}}{H'},\nonumber\\
B &=& \frac{q}{n}\, H'^{ - 1} d x^{(p - 1)}\wedge d x^{(p)},
\label{eq:conf}
\end{eqnarray}
where the string coupling $g_s = e^{\phi_0}$ with $\phi_0$ the
 asymptotic value of the dilaton, $i = p + 1, \cdots, 9$, 
$q$ and $n$ are two integers defined earlier and 
$H, H'$ are two harmonic functions
\begin{equation}
H = 1 + \frac{Q_p}{r^{7 - p}},\qquad H' = 1 + 
\frac{n^2}{n^2 + q^2} \frac{Q_p}{r^{7 - p}},
\label{eq:hf}
\end{equation}
where $r = \sqrt{y^i y^i}$ and $Q_p = g_s c_p \sqrt{n^2 + q^2}
 \alpha'^{(7 - p)/2}$ with $c_p = 2^{5 - p} \pi^{(5 - p)/2} \Gamma ((7 -
 p)/2)$.     
 
 In the NCYM decoupling
limit (see \cite{malr}, for example), we have,  
\begin{equation}
\alpha' \rightarrow 0,\quad g_s = \alpha'^{(5 - p)/2} \tilde g, \quad
q/n = \frac{\tilde b}{\alpha'}.
\label{eq:dl}
\end{equation}
Also the following redefined\footnote{It appears that we have
introduced additional scales for ${\tilde x}_{p - 1}, {\tilde x}_p$. As 
discussed earlier, these two fixed quantities are actually related to
$\tilde b$ because the noncommutative relation 
$[{\tilde x}_{p - 1}, {\tilde x}_p] = i \tilde b$. the value of 
${\tilde x}_p$ (or ${\tilde x}_{p - 1}$) should correspond to the
quantum-mechanical average of the corresponding operator. It should be
on the order of ${\tilde b}^{1/2}$.}
$\tilde {x}_\mu$'s rather than the original
$x_\mu$'s are kept fixed
\begin{equation}
x_{0,1, \cdots, (p - 2)} = \tilde x_{0, 1, \cdots, (p - 2)},\qquad
x_{(p - 1), p} = \frac{\alpha'}{\tilde b} \tilde x_{(p - 1), p
}.
\end{equation}

One can check easily from Eq.(2.5) that the asymptotic values for the 
metric, the dilaton and the $B$-field with respect to the fixed 
$\tilde x$'s  
correspond to the closed string moduli of Seiberg and Witten \cite{seiw}
\footnote{This is a natural choice. But one could do 
differently as in \cite{liw} where an effective string tension is 
introduced and the $\alpha'$ is set to 1.}, respectively, if the
relation between our $B$ and the Seiberg and Witten's $B^{\rm SW}$, i.e,
$B = 2\pi \alpha' B^{\rm SW}$, is used. Using Eq.\ (\ref{eq:occr}), we can have
$G_s = \alpha'^{(3 - p)/2} \tilde g \tilde b$ as given earlier.
   
In the NCYM decoupling limit, any fixed $r \ne 0$ region will decouple from
the region defined by $r = \alpha' u$ with $u$ fixed which provides the 
gravity dual description of the (D(p$-$2), Dp) system. The vortex of q 
D(p - 2) branes located at ${\tilde x}_{p - 1} = {\tilde x}_p = 0$ will
decouple from the rest of identical vortices.  For convenience,
let us write the NCYM  decoupling limit collectively here:
\begin{eqnarray}
&&\alpha' \rightarrow 0,\quad  q/n  = \frac{\tilde b} {\alpha'},\quad
g_s = \alpha'^{(5 - p)/2} \tilde g,\nonumber\\
&& x_{0, 1, \cdots, (p - 2)} = \tilde x_{0, 1, \cdots, (p - 2)}, \quad
x_{(p - 1), p} = \frac{\alpha'}{\tilde b} \tilde x_{(p - 1), p},\quad
r = \alpha' u,
\label{eq:cdlone}
\end{eqnarray}

where $\tilde b, \tilde g, u, \tilde x_\mu$ remain fixed. For the reason
explained earlier, we always take $n = 1$ in the following discussion.

      With this limit, the NCYM gauge coupling is given as
\begin{equation}
g_{\rm YM}^2 = (2\pi)^{p - 2}\, {\tilde b} {\tilde g},
\end{equation}
which is fixed as expected. The NCYM effective gauge coupling is
$g^2_{\rm eff} \approx g_{\rm YM}^2 u^{p - 3}$. In this paper, we will
focus solely on the gravity dual description of NCYM, which is expected to
be in the region where the perturbative calculations in super Yang-Mills
cannot be trusted, i.e.,
\begin{equation}
g^2_{\rm eff} \gg 1 \, \Rightarrow \, \left\{\begin{array}{ll}
                                             u \ll (\tilde g \tilde
					     b)^{1/(3 - p)}, & p < 3\\
                                             u \gg (1/(\tilde g \tilde
					     b)^{1/(p - 3)}, & p >
					     3\end{array}
\right.
\label{eq:ul}
\end{equation}
As discussed earlier, the relevant parameter for NCYM is $\tilde b$. So
we will investigate the validity of the gravity dual description with
respect to this parameter while keeping $\tilde g$ fixed. 
                                        
With the limit Eq.\ (\ref{eq:cdlone}), the gravity description for 
(D(p$-$2), Dp) is
\begin{eqnarray}
d s^2 &=& \alpha' \left\{f^{-1} (u) u^2 \left[- d {\tilde x}_0^2 + d {\tilde
x}_1^2 + \cdots + d {\tilde x}_{(p - 2)}^2 + \hat h \left( d {\tilde
x}_{(p - 1)}^2 + d {\tilde x}_p^2\right)\right] \right.\nonumber\\
&\,&\qquad \qquad \qquad \left. + f (u) \left(\frac{d
u^2}{u^2} + d \Omega_{8 - p}^2\right)\right\},\nonumber\\
e^{2 \phi} &=& {\hat g}^2 \frac{(a u)^{(7 - p) (p - 3)/2}}{1 + (a u)^{7
- p}},\nonumber\\
B &=& \frac{\alpha'}{\tilde b} \frac{(a u)^{7 - p}}{1 + (a u)^{7 - p}}
d {\tilde x}^{p - 1} \wedge d {\tilde x}^p,
\label{eq:gd}
\end{eqnarray}
where $\hat g = \tilde g {\tilde b}^{(5 - p)/2}$, $a = {\tilde
b}^{1/2}/(c_p \hat g)^{1/(7 - p)}$ with constant $c_p$ defined earlier,
and functions $f (u)$ and $\hat h$
are
\begin{equation}
f (u) = \frac {(c_p \hat g)^{2/(7 - p)}} { (a u)^{(3 - p)/2}},\qquad
\hat h = \frac{1}{1 + (a u)^{7 - p}}.
\label{eq:fh}
\end{equation}

 Note that the NCYM effective string coupling, $e^\phi$, is finite in the 
decoupling limit (since $\tilde b, \tilde g$ and $u$ remain fixed). The 
curvature in string units is
\begin{equation}
\alpha' R \approx \frac{1}{g_{\rm eff}} \sim \sqrt{\frac{u^{3 -
p}}{\tilde g \tilde b}} \sim \sqrt{\frac{(a u)^{3 - p}}{{\hat g}^{4/(7 -
p)}}}.
\label{eq:cv}
\end{equation}

From the NC field theory perspective, $u$ is an energy scale. The limit 
$u \rightarrow \infty$ means going to UV in the field theory.  
In this limit, unlike the cases for simple Dp branes without $B$-field 
analysed in \cite{itzmsy}, the effective string coupling, 
$e^\phi \sim (a u)^{(7 - p)(p - 5)/2}$, vanishes for $p < 5$ while 
the NCYM theory becomes UV free still for $p < 3$ (note that we have
$ 2 \le p \le 6$ in this paper). For $p > 3$,
the field theory breaks down and we need new degrees of freedom there.
However, from the gravity side for $ 5 > p > 3$, both the effective string
coupling and the curvature (see the second equation in (2.12) and Eq.(2.14)) 
vanishes and therefore
the gravity description is perfectly good. For $p = 5$, the curvature
still vanishes but the effective string coupling remains finite as
$e^\phi = \hat g$. Finally for $p = 6$, the string effective coupling blows up
and the dual description is needed. As pointed out in
\cite{aliosj}, 
even for this case, the worldvolume theory with $B$-field differs from that
without $B$-field in that the former decouples from gravity while the
latter does not \cite{ahagmoo,aliosj} in their respective decoupling
limits. The 
reason for this is now clear that the D6 branes with constant $B$-field 
are equivalent to a
system of infinitely many (i.e., q) D4 branes  
without $B$-field in the decoupling limit. The decoupling of gravity in
the latter system  
must imply the same in the former one.

      Let us now consider the gravity solutions of D(p$-$2) branes with
two additional isometries for $2 \le p \le 6$. They can be obtained from
(D(p$-$2), Dp) simply by setting the charge associated with Dp
branes to zero. In doing so, we need to use the original $B$-field, 
(rather than the
one given above where a gauge choice is made for NCYM) and we find that
as $n=0$, $H' =
1$ and so $B=0$. 
If we take the same limit as given in Eq.\ (\ref{eq:cdlone})
except for $q/n$ which is now replaced by $q = { \tilde
b}/\alpha'$
we
end up 
with\footnote{If we do not take $n = 1$ in the discussion of (D(p
$-$ 2), Dp) system above, we simply choose the $\tilde b$ here 
as $n$ times 
$\tilde b$ there.}, 
\begin{eqnarray}
d s^2 &=& \alpha' \left\{f^{- 1} (u) u^2 \left[- d {\tilde x}_0^2 + d {\tilde
x}_1^2 + \cdots + d {\tilde x}_{(p - 2)}^2 + \frac{1}{(au)^{7 - p}} 
\left( d {\tilde x}_{(p - 1)}^2 + d {\tilde x}_p^2\right)\right]\right.
\nonumber\\
&\,&\qquad\qquad \left. + 
f (u) \left(\frac{d
u^2}{u^2} + d \Omega_{8 - p}^2\right)\right\},\nonumber\\
e^{2 \phi} &=& {\hat g}^2 (a u)^{(7 - p) (p - 5)/2},\nonumber\\
B &=& 0,
\label{eq:dp2}
\end{eqnarray}
where $a, \hat g$ and $f(u)$ are the same as given before. As discussed
before, the above configuration should be explained as the one
describing $q \rightarrow \infty$ D(p - 2) branes located at 
${\tilde x}_{p - 1} = {\tilde x}_p = 0$.

Comparing Eq.\ (\ref{eq:dp2}) for $q$ D(p$-$2) branes  with
  Eq.\ (\ref{eq:gd}) for (D(p$-$2), Dp), we see that the
two configurations become the same, except for the $B$-field, if $au \gg
1$. The former has 
${\tilde B}_{(p - 1)p} = 0$ while the latter has a constant 
${\tilde B}_{(p - 1) p} = \alpha'/\tilde b$ (Note that the ${\tilde
B}_{(p - 1) p}$ is defined with respect to the fixed $\tilde x$
coordinates though the corresponding 2-form is inert under the
rescalings of coordinates). For the gravity configuration of
$q$ D(p$-$2) branes, any non-vanishing
constant $B$-field along directions transverse to the D(p$-$2) branes
can always be gauged away \cite{seiw}. So we conclude that the two 
configurations are identical if $a u \gg 1$. Further, as we will
  demonstrate, $a u \gg 1$ implies ${\tilde b}^{1/2} u \gg 1$, i.e., 
in the region of strong noncommutative effects. 
 Therefore
the crucial point for us to establish the equivalence between 
the $q$ D(p$-$2) branes  and Dp branes is to show whenever the
gravity description is valid, we always end up with $au \gg 1$. As we
will show in the following sections, this is indeed true. 

     Note also that the (D(p$-$2), Dp) system preserves the same number
(sixteen) of supersymmetries as D(p$-$2) branes with two additional
isometries. This is necessary for the equivalence. 
    
\section{The (D1, D3) system\\} 
  
	The decoupling limit for this system with classical 
$B_\infty = \tan \theta$ has been studied in \cite{malr}. Here we 
revisit this system with quantized  $B_\infty = \tan \theta = q/n$.
As explained in the previous sections, we shall take $n = 1$. Taking $p
= 3$ in the NCYM decoupling limit Eq.\ (\ref{eq:cdlone}), we have
\begin{eqnarray}
&&\alpha' \rightarrow 0,\quad  q  = \frac{\tilde b} {\alpha'},\quad
g_s = \alpha' \tilde g,\nonumber\\
&& x_{0, 1} = \tilde x_{0, 1}, \quad
x_{2, 3} = \frac{\alpha'}{\tilde b} \tilde x_{2, 3},\quad
r = \alpha' u,
\label{eq:cdltwo}
\end{eqnarray}
where $\tilde b, \tilde g, u, \tilde x_\mu$ remain fixed. In this limit,
the asymptotic region decouples from the near-horizon region which
describes the (D1, D3) system.     
  
   In particular, $q \gg n$ implies that D-strings in the system plays
   the dominating role over the $n$ D3 branes.
 Since the noncommutative geometry in 
${\tilde x}^2 {\tilde x}^3$ directions is controlled by the parameter 
$\tilde b$ which is in turn controlled by the large integer $q$,
therefore the non-commutativity is due to the infinitely many D-strings 
in the system. Notice that the open string coupling $G_s =
   \hat g$ with $\hat g = \tilde g \tilde b$, the closed string coupling
   at the IR\footnote{The open string coupling or the gauge coupling is
   also related to the closed string coupling at IR for the usual
 $AdS_5/CFT_4$ correspondence with fixed $q$ and $n$ as pointed out in 
\cite{lurthree}.} while the closed string
   coupling $g_s = \alpha' \tilde g$. This tells us that from the
   noncommutative open
   string perspective (or the noncommutative field theory description) 
the system looks
 like D3 branes while from the closed string side (or the gravity
   description) it behaves like D-strings if we compare the scalings of
these string couplings with those for simple D-branes 
(i.e., without $B$-field) given in
\cite{itzmsy} in their respective decoupling limits. 
 
	In this section, using the gravity dual description of NCYM,  
we will provide evidence for the claim that the (D1, D3) system is
reduced to a system of $q$ D-strings without $B$-field.   

     The gravity description for (D1, D3)  has the form
\begin{eqnarray}
ds^2 &=& \alpha' \left\{f^{- 1} u^2 \left[- d {\tilde x}_0^2 + d
{\tilde x}_1^2 
+  {\hat h} \left( d {\tilde x}_2^2 + d {\tilde x}_3^2\right)\right] + 
 f \left(\frac{d u^2}{u^2} + 
d \Omega_5^2\right) \right\},\nonumber\\
e^{2\phi} &= & \hat{g}^2\, \hat h,\nonumber\\
\tilde{B}_{23} &=& \frac{\alpha'}{\tilde b}\frac{a^4 u^4}{1 + a^4 u^4},
\quad A_{01} = 
\alpha'\frac{\tilde b}{\hat g} \frac{u^4}{4\pi \hat g},
\quad \tilde{H}_{0123u} = \alpha'^2 
\frac{\hat h}{\hat g  f^2}\partial_u \left(u^4\right),\nonumber\\
\hat h &=& \frac{1}{1 + a^4 u^4}, \quad a^2 = \frac{\tilde b}{ f}, 
\quad f^2 = 4\pi \hat g.
\label{eq:d1d3gd}
\end{eqnarray}
This has been obtained by setting $p = 3$ in Eq.\ (\ref{eq:gd}) and
we have also included in the above 
the RR 2-form $A$ associated with the D-strings and the self-dual
RR 5-form field strength $H$.  Note also that 
$\tilde{B}_{23}$, $A_{01}$ and $\tilde{H}_{0123u}$ are now defined with
respect to the new coordinates through the corresponding forms which are
inert under the rescalings of coordinates. 

The curvature in string units is given by Eq.\ (\ref{eq:cv}) and can
also be read off from the metric in (3.2) as
\begin{equation}
\alpha' R \approx \frac{1}{g_{\rm eff}} \sim \frac{1}{\sqrt{ \hat g}},
\label{eq:cv3}
\end{equation}
where $g_{\rm eff}^2 = 2\pi \hat g$ for $p = 3$. The gravity description
is good if the effective string coupling, $e^\phi$, and the above
curvature in string units are small, i.e., 
\begin{equation}
\frac{\hat g}{\sqrt{1 + (a u)^4}} \ll 1,\qquad g^2_{\rm eff} \sim \hat g \gg 1.
\label{eq:gvl3}
\end{equation}
As usual, the validity of gravity description requires large effective
gauge coupling $g_{\rm eff}$ which implies that the field theory
description breaks down. The above equation says 
\begin{equation}
au \gg {\hat g}^{1/2} \gg 1.
\label{eq:gl3}
\end{equation}
As explained in the previous section, we always keep $\tilde g$
fixed. So large $\hat g = \tilde g \tilde b$ means large $\tilde b$
which is consistent with the picture that the non-commutative effect
is important for large $au$ in the gravity description.  

   Notice that since $a \sim {\tilde b}^{1/4}/{\tilde g}^{1/4}$ is large with
large $\tilde b$ and fixed $\tilde g$, large $a u$ does not
necessarily mean large $u$. Therefore the condition (3.5) for $u$ does not mean
that we approach the boundary which is at $u = \infty$. Further we have 
${\tilde b}^{1/2} u \gg a u \gg 1$ which implies that the dynamical
degrees of freedom are due to the D-strings as explained before.
Let us examine
the configuration \ (\ref{eq:d1d3gd}) under Eq.\ (\ref{eq:gl3}) with
fixed $\tilde g$. In this case, $\hat h \sim 1/(au)^4$ and ${\tilde
B}_{23}$ approaches its constant but small value $\frac{\alpha'} {\tilde
b}$ for large $\tilde b$ even if it is measured in terms of
$\alpha'$. The 5-form field strength 
${\tilde H}_{0123u} /\alpha'^2 \sim 1 / [({\tilde
g}^{5/4} {\tilde b}^{11/4}) au]$ approaches zero in the limit. 
However, via 
Eq.\ (\ref{eq:gl3}), the RR
2-form $A_{01}/\alpha' \sim {\tilde b} u^4 /{\hat g}^2 = 
\tilde g (a u)^4/{\hat g}^2 \gg 1$ survives. So everything goes
over to
that of q D-strings as given in 
Eq.\ (\ref{eq:dp2}) (with $p = 3$) in the same limit Eq.\ (\ref{eq:cdltwo}).
In other words, when the gravity dual description of NCYM for D3 branes
with large asymptotic $B$-field is valid, this gravity system is reduced
to q D-strings without B-field. 

In the region ${\hat g}^{1/2} \gg au \gg 1$,  the effective string
 coupling is large.   So, 
we need to go to the S-dual gravity description.   
In this case, (F, D3) is reduced to q F-strings without RR 2-form field 
since under the
 S-duality, D-strings become F-strings ((D1, D3) becomes (F, D3)).
Under S-duality the NS 2-form $B$ becomes a RR
 2-form while the RR 2-form $A$ becomes a NS 2-form and this is the reason
for the appearance of F-strings. 

 Under S-duality, we have
\begin{eqnarray}
&&l_s^2 \rightarrow {l'}_s^2 \equiv g_s l_s^2,\qquad g_s \rightarrow 
{g'}_s \equiv \frac{1}{g_s},\nonumber\\
&&e^\phi \rightarrow e^{\phi'} \equiv e^{-\phi}, \qquad 
d s^2 \rightarrow d {s'}^2 \equiv g_s e^{- \phi} d s^2,
\label{eq:sdt}
\end{eqnarray}
where $l_s = \sqrt{\alpha'}$ is the string length scale in the original
variables. So, the new metric and the dilaton are given as,
\begin{eqnarray}
d {s'}^2 &=& H'^{- 1/2}\left[ \frac{H'}{H} \left( - d x_0^2 + d
x_1^2\right) + d x_2^2 + d x_3^2\right] + H'^{1/2} dy^i d
y^i,\nonumber\\
e^{2 \phi'} & = & {g'}_s^2 \frac{H'}{H},
\label{eq:fd3}
\end{eqnarray}
where the harmonic functions continue to be given by Eq.\ (\ref{eq:hf})
but  $g_s$ and $\alpha'$ in the charge $Q_3$ are replaced by 
their S-dual values through the relations in Eq.\ (\ref{eq:sdt}).
One can check that the metric and the dilaton are indeed those
describing the (F, D3) system as given in \cite{lurtwo}.

Following \cite{aliosj}, the decoupling limit for this system is the
same as those given in Eq.\ (\ref{eq:cdltwo}). The only difference is that
we express them in terms of ${l'}_s$ and ${g'}_s$ through $g_s =
1/{g'}_s$ and $\alpha' = {g'}_s {l'}_s^2 \rightarrow 0$. In the
decoupling limit, we have the following for (F, D3)
\begin{eqnarray}
d {s'}^2 &=& {l'}_s^2 {\hat h}^{- 1/2}\left\{ 
\frac{(\hat{g'} u)^2}{\sqrt{4\pi \hat{g'}}} \left[- d {\tilde x}_0^2
+ d {\tilde x}_1^2 + \hat h \left(d {\tilde x}_2^2 + d {\tilde
x}_3^2\right) \right] + \sqrt{4 \pi \hat{g'}}\left(\frac{d u^2}{u^2} + 
d \Omega_5^2\right)\right\},\nonumber\\
e^{2 \phi'} &=& {\hat{g'}}^2 \left( 1 + (au)^4\right),
\label{eq:gdfd3}
\end{eqnarray}
where $\hat h$ is given in Eq.\ (\ref{eq:hf}) with the same
definition for the parameter $a$  there and $\hat{g'} =
1/{\hat g} = 1/{\tilde g \tilde b}$. The curvature measured in ${l'}_s$
is given as,
\begin{equation}
{l'}_s^2 R \sim \frac{1}{\sqrt{\hat{g'} ( 1 + a^4 u^4)}}.
\label{eq:cvfd3}
\end{equation}
 
The gravity description is good when the effective string coupling, 
$e^{\phi'}$, and the curvature are small, i.e.,
\begin{equation}
\frac{1}{\hat{g'}^{1/4}} \ll au \ll \frac{1}{\hat{g'}^{1/2}},
\label{eq:gvdfd3}
\end{equation}
which is true if $\hat{g'} = 1/{\hat g} \ll 1$. In the original $\hat
g$, Eq.(3.10) says  $ 1 \ll \hat{g}^{1/4} \ll au \ll \hat{g}^{1/2}$. For
fixed $\tilde g$, $\hat g \gg 1$ implies large $\tilde b$ and $au \gg 1$
implies $\hat h \sim 1/(au)^4$. We have again ${\tilde b}^{1/2} u
\gg au \gg 1$. Then the gravity description for (F, D3)
goes over to that for q F-strings  in the
similar limit as expected.

\section{The remaining (D(p$-$2), Dp) systems}

In this section we discuss the equivalence for the rest of the (D(p$-$2), Dp)
systems.

\noindent{\bf (D0, D2):}

The decoupling limit for this system is
\begin{eqnarray}
&&\alpha' \rightarrow 0,\quad q = \frac{\tilde b}{\alpha'},\quad 
g_s = \alpha'^{3/2} \tilde g,\nonumber\\
&& x_0 = \tilde{x}_0,\quad x_{1, 2} = \frac{\alpha'}{\tilde b}\,
\tilde{x}_{1, 2},\quad r = \alpha' u,
\label{eq:cdlthree}
\end{eqnarray}
where $\tilde b, \tilde g, u, \tilde{x}_\mu$ remain fixed. Under this
limit, the gravity description for (D0, D2) is
\begin{eqnarray}
d s^2 &=& \alpha' \left\{f^{-1} (u) u^2 \left[- d {\tilde x}_0^2  + \hat h 
\left( d {\tilde
x}_1^2 + d {\tilde x}_2^2\right)\right] + f (u) \left(\frac{d
u^2}{u^2} + d \Omega_6^2\right)\right\},\nonumber\\
e^{2 \phi} &=& {\hat g}^2 \frac{(a u)^{- 5/2}}{1 + (a u)^5},\nonumber\\
B &=& \frac{\alpha'}{\tilde b} \frac{(a u)^5}{1 + (a u)^5}
d {\tilde x}^1 \wedge d {\tilde x}^2,\nonumber\\
f (u) &=& (a u)^{- 1/2} (c_2 \hat g)^{2/5}, \quad \hat h  =
\frac{1}{1 + (a u)^5},
\label{eq:gd2}
\end{eqnarray}
where $\hat g = \tilde g \tilde{b}^{3/2}$, $a = {\tilde b}^{1/2}/(6 \pi^2
\hat g)^{1/5}$. The curvature in string units is
\begin{equation}
\alpha' R \approx \frac{1}{g_{\rm eff}} \sim \sqrt{\frac{u}{\tilde g
\tilde b}} \sim \sqrt{\frac{a u}{{\hat g}^{4/5}}}.
\label{eq:cv2}
\end{equation}
The gravity description is valid if the effective string coupling,
$e^\phi$, and the above curvature in string units are small, i.e.,
\begin{equation}
{\hat g}^{4/15} \ll au \ll {\hat g}^{4/5},
\label{eq:gvl2}
\end{equation}
which can hold only if $\hat g \gg 1$. For fixed $\tilde g$, this
implies large $\tilde b$ and therefore the large non-commutativity. We
also have ${\tilde b}^{1/2} u \gg au \gg 1$ which implies that the
dynamical degrees of freedom are due to D0 branes and the noncommutative
effect is important. The $au \gg 1$ implies  $\hat h \sim 1/(au)^5$. 
Since $a \sim
{\tilde b}^{1/5}/{\tilde g}^{1/5}$, this is also large.  So large 
$au$ does not
necessarily mean large $u$. With the above, we have the metric 
and the dilaton reduced to those of q D0 branes 
in the same limit as given in Eq.\ (\ref{eq:dp2}) for $p = 2$.
Note that $B_{12}/\alpha' = 1/{\tilde b}$ is now a small constant
 for large $\tilde b$. 

   In the region $a u \ll {\hat g}^{4/15}$, the effective string
   coupling, 
$e^\phi$, becomes large. So, we need to lift this system to eleven dimensions.
It can still be described by the 11D supergravity if the curvature in
11D Plank units remains small.  We have the 11D curvature in 11D Plank
units as
\begin{equation}
l_p^2 R \sim e^{2\phi/3} \frac{1}{g_{\rm eff}} \sim
   \frac{\hat{g}^{4/15}}{(a u)^{1/3} (1 + a^5 u^5)^{1/3}} \ll 1
\label{eq:11cv2}
\end{equation}
which gives $ au \gg \hat{g}^{2/15}$. So the 11D gravity description for
the system is valid if $ \hat{g}^{2/15} \ll au \ll \hat{g}^{4/15}$. For
this to be true, we must have $\hat {g} \gg 1$, therefore  $\tilde
b$ becomes large for fixed $\tilde g$. Precisely with these, 
the gravity description
for the lifted (D0, D2) branes is the same as that for the lifted D0
branes. 

\bigskip
\noindent{\bf (D2, D4):}

	  The decoupling limit for this system is	  
\begin{eqnarray}
&&\alpha' \rightarrow 0,\quad  q  = \frac{\tilde b} {\alpha'},\quad
g_s = \alpha'^{1/2} \tilde g,\nonumber\\
&& x_{0, 1, 2} = \tilde x_{0, 1, 2}, \quad
x_{3, 4} = \frac{\alpha'}{\tilde b} \tilde x_{3, 4},\quad
r = \alpha' u,
\label{eq:cdlf}
\end{eqnarray}       
where $\tilde b, \tilde g, u, \tilde{x}_\mu$ remain fixed. Under this
limit, the gravity description for (D2, D4) is
\begin{eqnarray}
d s^2 &=& \alpha' \left\{f^{-1} (u) u^2 \left[- d {\tilde x}_0^2 + d {\tilde
x}_1^2 + d {\tilde x}_2^2 + \hat h \left( d {\tilde
x}_3^2 + d {\tilde x}_4^2\right)\right] + f (u) \left(\frac{d
u^2}{u^2} + d \Omega_4^2\right)\right\},\nonumber\\
e^{2 \phi} &=& {\hat g}^2 \frac{(a u)^{3/2}}{1 + (a u)^3},\nonumber\\
B &=& \frac{\alpha'}{\tilde b} \frac{(a u)^3}{1 + (a u)^3}
d {\tilde x}^3 \wedge d {\tilde x}^4,\nonumber\\
f (u) &=& \left(\pi \hat g\right)^{2/3} (a u)^{1/2},\qquad \hat h =
\frac{1}{1 + (a u)^3},
\label{eq:gd4}
\end{eqnarray}
where $\hat g = \tilde g {\tilde b}^{1/2}$, $a = {\tilde
b}^{1/2}/(\pi \hat g)^{1/3}$. The curvature in string units is
\begin{equation}
\alpha' R \approx \frac{1}{g_{\rm eff}} \sim \frac{1}{\sqrt{\tilde g
\tilde b u}} \sim \frac{1}{(au)^{1/2} {\hat g}^{2/3}}.
\label{eq:cv4}
\end{equation}

Unlike the case for D4 branes without $B$-field, the closed string
coupling $g_s = \alpha'^{1/2} \tilde g$ vanishes (rather than blows up)
in the decoupling limit. It is still proper to consider the 10D gravity
description for this system. In order to have a valid gravity
description, the effective string coupling, $e^\phi$, and the above
curvature in string units need to remain small, i.e.,
\begin{equation}
\hat{g}^2 \, \frac{(a u)^{3/2}}{1 + (a u)^3} \ll 1,\qquad 
a u \gg 1/\hat{g}^{4/3}.
\label{eq:gvd4}
\end{equation}
We have two cases here depending on whether $\hat g \ge 1$ or $\hat g <
1$. For $\hat g < 1$, the first equation in (4.9) can be satisfied 
if $a u \ll 1$
or $a u \gg \hat{g}^{4/3}$. However, $au \ll 1$ is  imcompatible with the
second equation in (4.9) if $\hat g < 1$. So we have $a u \gg
1/\hat{g}^{4/3} > 1$ for which the gravity description is valid. In this
case, since $\hat g < 1$, the $\tilde b$ is not large and $a$ is not
large, either. Therefore  $au \gg 1$ does require large $u$.
So, when the gravity description is valid, we are approaching the boundary.
Similar analysis tells us that $au \gg \hat{g}^{4/3} > 1$ is the
condition for a valid gravity description if $\hat g > 1$. Again,  
we need to approach the boundary, i.e., large $u$ to validate the
gravity description. The crucial thing here is that $au \gg 1$ and 
${\tilde b}^{1/2} u \gg 1$ hold
true whenever the gravity description is valid. Now we have $\hat h \sim
1/(au)^3$. The metric and the dilaton go over to those of q D2 branes
 in the same limit. But there is a
difference here from the previous cases considered, i.e., the $B$-field
no longer remains small the way it happens before. It actually becomes 
a constant, i.e.,
$\tilde{B}_{34}/\alpha' = 1/{\tilde b}$. So long as the gravity description
for the D2 branes is concerned, a constant
${\tilde B}_{23}$ makes no difference from zero $\tilde {B}_{23}$. 
Therefore, we expect that the (D2, D4) system is reduced to q D2 branes.   

  For $\hat g < 1$, the effective string coupling can never be
  large. However, for $\hat g > 1$, the
effective string coupling, $e^\phi$, is large in the region 
$au \ll \hat{g}^{4/3}$. In that case, we need 
to lift (D2, D4)
to eleven dimensions. The 11D gravity description is valid if the curvature
in 11D Planck units is small, i.e.,
\begin{equation}
l_p^2 R \sim e^{2\phi/3} \frac{1}{g_{\rm eff}} \sim \frac{1}{
(1 + a^3 u^3)^{1/3}} \ll 1,
\end{equation}
which requires $au \gg 1$. So the lifted (D2, D4) can be described by
  11D gravity if $1 \ll au \ll \hat{g}^{4/3}$. This can be true only
if $\hat g \gg 1$. For fixed $\tilde g$, it means large $\tilde b$,
  therefore large $a \sim \tilde{b}^{1/3}/ \tilde{g}^{1/3}$. So, we do not
  necessarily have large $u$ here. For this description, the lifted
(D2, D4) goes over to the lifted D2 branes. 

\bigskip
\noindent{\bf (D3, D5):}

	   The corresponding decoupling limit is
\begin{eqnarray}
&&\alpha' \rightarrow 0,\quad  q  = \frac{\tilde b} {\alpha'},\quad
g_s = \tilde g,\nonumber\\
&& x_{0, 1, 2, 3} = \tilde x_{0, 1, 2, 3}, \quad
x_{4, 5} = \frac{\alpha'}{\tilde b} \tilde x_{4, 5},\quad
r = \alpha' u,
\label{eq:cdlfive}
\end{eqnarray}
where $\tilde b, \tilde g, u, \tilde x_\mu$ remain fixed.
With this limit, the gravity description of (D3, D5) is
\begin{eqnarray}
d s^2 &=& \alpha' \left\{f^{-1} (u) u^2 \left[- d {\tilde x}_0^2 + d {\tilde
x}_1^2 + d {\tilde x}_2^2 + d {\tilde x}_3^2 + \hat h \left( d {\tilde
x}_4^2 + d {\tilde x}_5^2\right)\right] + f (u) \left(\frac{d
u^2}{u^2} + d \Omega_3^2\right)\right\},\nonumber\\
e^{2 \phi} &=& {\hat g}^2 \frac{(au)^2}{1 + (a u)^2},\nonumber\\
B &=& \frac{\alpha'}{\tilde b} \frac{(a u)^2}{1 + (a u)^2}
d {\tilde x}^4 \wedge d {\tilde x}^5,\nonumber\\
f (u) &=& \hat g (a u),\qquad \hat h = \frac{1}{1 + (a u)^2},
\label{eq:gd5}
\end{eqnarray}
where $\hat g = \tilde g$, $a = {\tilde
b}^{1/2}/( \hat g)^{1/2}$.  The curvature in string units is
\begin{equation}
\alpha' R \approx \frac{1}{g_{\rm eff}} \sim \frac{1}{
\sqrt{\tilde g \tilde b u^2}} \sim \frac{1}{(au)\hat g}.
\end{equation}
Again unlike the case for D5 branes without $B$-field, the string coupling
$g_s = \tilde g$ is fixed (rather than blows up). It resembles that for
D3 branes without $B$-field in the decoupling limit. The gravity
description is good if both the effective string coupling, $e^\phi$, and
the above curvature in string units remain small. This requires
\begin{equation}
a u \gg \frac{1}{\hat g},  \qquad \hat g < 1.
\end{equation}
{}From above we have $au \gg 1$ and ${\tilde b}^{1/2} u \gg 1$. 
Then both the dilaton and the $B$-field become 
constant. We also have $\hat h \sim 1/(au)^2$ now. We do not have any
condition for $\tilde b$ which implies that $u$ itself must be large,
i.e. we need to approach the boundary. The gravity 
description goes over to that for q D3 branes in the same decoupling limit.

    When $\hat g \gg 1$, we need to go to its S-dual description. 
The $B$-field becomes RR 2-form field. The D5 branes become NS 5-branes.
So, (D3, D5) goes over to (D3, NS5).
Using Eq.\ (\ref{eq:sdt}), we have
\begin{eqnarray}
d s'^2 &=& {l'}^2_s \hat{h}^{- 1/2}\left\{\frac{(\hat{g'} u)^2}{a u}
 \left[- d {\tilde x}_0^2 + d {\tilde
x}_1^2 + d {\tilde x}_2^2 + d {\tilde x}_3^2 + \hat h \left( d {\tilde
x}_4^2 + d {\tilde x}_5^2\right)\right] + au \left(\frac{d
u^2}{u^2} + d \Omega_3^2\right)\right\},\nonumber\\
e^{2 \phi'} &=& \hat{g'}^2 ( 1 + a^2 u^2),
\label{eq:gdd5}
\end{eqnarray}
where $\hat{g'} = 1/\hat g$, $\hat h$ continues to be the one given in 
Eq.\ (\ref{eq:gd5}) and the same is for the parameter $a$. The curvature
in dual string units is
\begin{equation}
{l'}^2_s R = \frac{1}{au (1 + a^2 u^2)^{1/2}}.
\end{equation}
The dual gravity description is valid when we have
$1 \ll au \ll 1/\hat{g'} = \hat g$ which ensures that both the effective
string coupling and the curvature in the dual string units remain small. 
With this, $\hat h \sim
1/(au)^2$. Therefore, the (D3, NS5) goes over to the S-dual D3 branes.
 Note also that $\tilde b$ does not
play any role
here and we expect that we need to approach the boundary. This is indeed
true from $au \gg 1 \rightarrow \tilde {b}^{1/2} u \gg \hat{g}^{1/2} \gg
1$.   

\bigskip
\noindent{\bf (D4, D6):}

	  This is our last system. The decoupling limit is
\begin{eqnarray}
&&\alpha' \rightarrow 0,\quad  q  = \frac{\tilde b} {\alpha'},\quad
g_s = \alpha'^{- 1/2} \tilde g,\nonumber\\
&& x_{0, 1, \cdots, 4} = \tilde x_{0, 1, \cdots, 4}, \quad
x_{5, 6} = \frac{\alpha'}{\tilde b} \tilde x_{5, 6},\quad
r = \alpha' u,
\label{eq:cdlsix}
\end{eqnarray}
where $\tilde b, \tilde g, u, \tilde x_\mu$ remain fixed. 
Note that this system differs from the previous ones in that the
string coupling $g_s$ blows up in the decoupling limit. But it blows up
the same way as that for D4 branes without $B$-field rather than that
for D6 branes without $B$-field. Because of this blowing up, the theory
is better analysed in eleven dimensions.

   For the gravity description, the effective
string coupling, $e^\phi$, is our real concern. The type IIA gravity  
description of (D4, D6) in the above limit is
\begin{eqnarray}
d s^2 &=& \alpha' \left\{f^{-1} (u) u^2 \left[- d {\tilde x}_0^2 + 
d {\tilde x}_1^2 + \cdots + d {\tilde x}_4^2 + \hat h \left( d {\tilde
x}_5^2 + d {\tilde x}_6^2\right)\right] + f (u) \left(\frac{d
u^2}{u^2} + d \Omega_2^2\right)\right\},\nonumber\\
e^{2 \phi} &=& {\hat g}^2 \frac{(a u)^{3/2}}{1 + a u},\nonumber\\
B &=& \frac{\alpha'}{\tilde b} \frac{a u}{1 + a u}
d {\tilde x}^5 \wedge d {\tilde x}^6,\nonumber\\
f (u) &=& (\hat g /2)^2 (au)^{3/2},\qquad \hat h = \frac{1}{1 + a u},
\label{eq:gd6}
\end{eqnarray}
where $\hat g = \tilde g {\tilde b}^{- 1/2}$, $a = {\tilde
b}^{1/2}/( \hat g/2)$. The curvature in string units is
\begin{equation}
\alpha' R \approx \frac{1}{g_{\rm eff}}\sim \frac{1}{
\sqrt{\tilde g \tilde b u^3}} \sim \frac{1}{{\hat g}^2 (a u)^{3/2}}.
\label{eq:cv6}
\end{equation}
The gravity description is valid if both the effective string coupling
and the above curvature in string units are small. This gives 
$1/\hat{g}^{4/3} \ll au \ll 1/\hat{g}^4$ which can hold true only if
$\hat g \ll 1$.  For fixed $\tilde g$, this implies large $\tilde b$.
The above also implies $au \gg 1/\hat{g}^{4/3}\gg 1$ and 
${\tilde b}^{1/2} u \gg 1/\hat{g}^{1/3} \gg 1$. Then $\hat h \sim 1/(au)$.
Here ${\tilde B}_{56}/\alpha' = 1/\tilde b$
vanishes for large $\tilde b$. So 
the (D4, D6) system goes over to q D4 branes in the same limit. 

   In the region $au \gg 1/\hat{g}^4 \gg 1$, the effective string coupling is 
large. We need to lift the (D4, D6) system to eleven dimensions. The 11D
gravity description is valid if the following curvature
in 11D Planck units is small, i.e., 
\begin{equation}
l_p^2 R \sim e^{2\phi/3}\frac{1}{g_{\rm eff}} \sim \frac{1}{\hat{g}^{4/3}
au (1 + au)^{1/3}} \ll 1,
\end{equation}
which gives $au \gg 1/\hat g$.  So $a u \gg 1/\hat{g}^4 \gg 1$ already 
ensures small curvature in 11D Planck units. Since we have $au \gg 1$,
${\tilde b}^{1/2} u \gg 1/\hat{g}^3 \gg 1$
and $\hat g \ll 1$, the lifted (D4, D6) goes over to the lifted D4
branes. 
  
\section{Conclusion}    
In this paper, we have investigated what causes the noncommutativity in
terms of gravity dual description of NCYM for systems of non-threshold
bound states (D(p$-$2), Dp)
with $2 \le p \le 6$ in type II string theories.  Our study along with 
previous works strongly indicates that in the NCYM decoupling limit, the
 Dp branes with a constant $B$-field represent dynamically a system of 
infinitely many D(p$-$2) branes 
without $B$-field. In this limit, the corresponding $(p + 1)$-dimensional
NCYM is just a better description of $(p - 1)$-dimensional ordinary
Yang-Mills with gauge group $U(q)$ with $q \rightarrow \infty$ in that
it 
has a finite gauge
group $U (n)$ and a finite gauge coupling. The price we pay is that we
have noncommutative space in the directions of nonvanishing
$B$-field and we need to introduce the $\ast$ product for functions.   
In other words, we find that the 
noncommutativity is entirely due
to our choice in employing Dp branes to describe the dynamics of 
infinitely many D(p$-$2) branes 
without $B$-field. We give a clear explanation about the new parameter
$\tilde b$ which is related to the intrinsic energy scale 
$1/{\tilde b}^{1/2}$
for the D(p$-$2) branes.

	In particular,  we have shown that
the gravity configuration of (D(p$-$2), Dp) system or Dp branes with
a constant $B$-field is reduced to that
of infinitely many D(p$-$2) branes without $B$-field in the 
NCYM limit and in the region of valid gravity 
description.  
The present study provides also the reason
that the D6 brane with $B$-field can decouple from the gravity while the
simple D6 brane without $B$-field cannot in their respective 
decoupling limits. We expect that D7 branes with a rank 2 $B$-field can
decouple from gravity, too, in the NCYM decoupling limit since it
describes infinitely many D5 branes with two additional isometries and 
without $B$-field. But this will not be true for D8 or D9 branes with a 
rank 2 $B$-field. Our study indicates that in general Dp branes with a
rank $r$ constant $B$-field, this system can decouple from gravity in the
NCYM limit if $p' = p - r < 6$. In other words, the higher the rank $r$
is, the better chance we have for the Dp branes to decouple from
gravity. For example, D8 branes with rank 4 $B$-field can decouple from 
gravity.

  \acknowledgments
We would like to thank Nobuyuki Ishibashi for an e-mail correspondence
and for bringing our  attention to reference \cite{oku}, and Ricardo Schiappa
for an e-mail correspondence. Especially, we would like to thank
R.-G. Cai and N. Ohta for a very fruitful e-mail correspondence 
which leads us to refine
some points in section 1 and 2. We would also like to
thank Mike Duff for
reading the manuscript, Jim Liu and Leopoldo Pando-Zayas for discussions. 
JXLU acknowledges the support of U. S. Department of Energy.

  \end{document}